\newif\ifeprint \eprinttrue                            %%
\ifeprint \pdfoutput=1                                 %%
\documentclass                                         %%
    [rmp,amsmath,twocolumn,floatfix,                   %%
     showkeys,letterpaper]{revtex4}                    %%
\usepackage{hyperref,orcidlink,newtxtext,newtxmath}    %%
\hypersetup{pdftitle=                                  %%
  {Geodesics on an arbitrary ellipsoid of revolution}, %%
  pdfauthor={C. F. F. Karney}}                         %%
\newcommand{\doi}[1]{\href                             %%
  {https://doi.org/#1}{doi:\discretionary{}{}{}#1}}    %%
\else                                                  %%
\documentclass[pdflatex,sn-basic]{sn-jnl}
\renewcommand{\S}{\textsection}
\catcode`\|=12 % not sure why | is active, make it an "other" character
\fi                                                    %%
\usepackage{array,graphicx}
\usepackage[T1]{fontenc}        % for Polish letters
\DeclareMathOperator*{\sump}{\sum\nolimits^\prime}
\DeclareMathOperator*{\Ar}{Ar}
\usepackage{breakurl}
\ifpdf\def\urlalt#1#2{\burlalt{#2}{#1}}\else\let\urlalt=\burlalt\fi
\def\dlmf#1#2{\urlalt{http://dlmf.nist.gov/#2}{#1}}

\renewcommand{\d}{\mathrm d}
\newcommand{\half}{{\textstyle\frac12}}
\def\abs#1{\lvert#1\rvert}
\def\figuredir{figures}
\begin{document}
\title{Geodesics on an arbitrary ellipsoid of revolution}
\ifeprint                                             %%
\setlength\tabcolsep{0.4em}                           %%
\author{Charles F. F. Karney\,\orcidlink              %%
  {0000-0002-5006-5836}}                              %%
\email[Email: ]{charles.karney@sri.com}               %%
\thanks{\href                                         %%
  {mailto:karney@alum.mit.edu}{karney@alum.mit.edu}.} %%
\affiliation{SRI International,                       %%
  Princeton, NJ 08540-6449, USA}                      %%
\date{\today}                                         %%
\else                                                 %%
\author{Charles F. F. Karney}
\email{charles.karney@sri.com}\email{karney@alum.mit.edu}
\affil{SRI International,
  Princeton, NJ 08540-6449, USA}
\equalcont{ORCiD: \href
  {https://orcid.org/0000-0002-5006-5836}
  {0000-0002-5006-5836}}
\fi                                                   %%

\ifeprint\begin{abstract}\else                        %%
\abstract{
\fi                                                   %%
The algorithms given in Karney, J.~Geodesy {\bf 87}, 43--55 (2013), to
compute geodesics on terrestrial ellipsoids are extended to apply to
ellipsoids of revolution with arbitrary eccentricity.  For the direct
and inverse geodesic problems, this entails implementing the formulation
in terms of elliptic integrals given by Legendre and Cayley.  The
integral for the area of geodesic polygons is computed in terms of the
discrete sine transform of the integrand.  In all cases, accuracy close
to full
machine precision is achieved.  An open-source implementation of these
algorithms is available.
\ifeprint\end{abstract}\else                 %%
}
\fi                                          %%

\keywords{geometrical geodesy, geodesics, eccentric ellipsoids,
polygonal areas, numerical methods}
\maketitle

\section*{List of symbols}

\begin{center}
\setlength{\extrarowheight}{1ex}
\begin{tabular*}{\columnwidth
}{@{\extracolsep{0.5em}}>{$}r<{$}l}
a & equatorial radius of the ellipsoid \\
b & its polar semi-axis \\
f & $(a-b)/a$, the flattening \\
n & $(a-b)/(a+b)$, the third flattening \\
e & $\sqrt{a^2-b^2}/a$, the eccentricity \\
e' & $\sqrt{a^2-b^2}/b$, the second eccentricity \\
\phi & geographic latitude, north positive \\
\lambda & longitude, east positive \\
\alpha & azimuth of the geodesic, clockwise from north \\
\alpha_0 & azimuth at the node \\
s & distance along the geodesic \\
\beta & parametric latitude; $a\tan\beta = b\tan\phi$ \\
\omega & longitude on the auxiliary sphere \\
\sigma & arc length on the auxiliary sphere \\
c & the authalic radius \\
S & the area between the geodesic and the equator\\[1ex]
\end{tabular*}
\end{center}
The quantities $s$, $\sigma$, $\lambda$, and $\omega$ are measured from
the node, the point where the geodesic crosses the equator in the
northward direction.  A single subscript refers to a specific point,
e.g., $s_1$ measures the distance from the node to point $1$; for $s$,
$\sigma$, and $\lambda$, a double subscript denotes a difference, e.g.,
$\lambda_{12} = \lambda_2 - \lambda_1$.

\section{Introduction}

In an earlier paper, {\it Algorithms for geodesics} \citep[henceforth
referred to as {\it AG}]{karney13}, I presented algorithms for solving
geodesic problems on an ellipsoid of revolution.  This built on the
classic work of \citet{bessel25-en} and \citet{helmert80-en} to give
full double precision accuracy for ellipsoids with flattening satisfying
$\abs f \le \frac1{100}$.  A small improvement, described in
Appendix~\ref{reverted-dist}, extended this range $\abs
f \le \frac1{50}$.  Perhaps the most consequential innovation of {\it
AG} was the reliable solution of the inverse geodesic problem,
computing the shortest path between two arbitrary points on the
ellipsoid.  (Previously, the state of the art was provided
by \citet{vincenty75a}, which sometimes fails to converge.)  Another
important advance was turning the formulation of \citet{danielsen89} for
the area between a geodesic segment and the equator into an algorithm
for computing accurately the area of any geodesic polygon, a polygon
whose edges are geodesics.  These algorithms allow geodesic problems to
be reliably solved and this, in turn, has meant that they have been
widely adopted in a variety of disciplines.

The present paper seeks to extend the treatment to arbitrary ellipsoid
(in this paper, the term ``ellipsoid'' should be understood to mean
``ellipsoid of revolution'').  The condition $\abs f \le \frac1{50}$
covers all terrestrial applications.  However, for the giant planets in
our solar system, $f$ lies outside this range (for Saturn, we have
$f\approx \frac1{10}$).  For geodesic calculations, the solutions of the
direct and inverse problems, this is a matter of realizing the
formulation in terms of elliptic integrals, originally given by
Legendre, as a usable algorithm.  Generalizing the computation of the
area of a geodesic polygon required a novel approach using the discrete
sine transform.

Although the mathematical techniques
used in this paper might be daunting for some readers, the
concepts of geodesics with their interlocking definitions in terms of
straightest lines and shortest paths are readily grasped.  Furthermore,
geodesics have well understood properties, for example:
\begin{itemize}
\ifeprint\itemsep-0.4ex plus 0.5ex \fi %%
\item
the shortest path from a point to a line intersects the line at right
angles;
\item
shortest geodesics satisfy the triangle inequality and the other
conditions that define a metric space;
\item
consequently, nearest neighbor problems can be solved efficiently using
a vantage-point tree \citep{uhlmann91,yianilos93}.
\end{itemize}
It's therefore possible, indeed it might be preferable, to tackle many
problems involving geodesics by treating the implementation described
here as a reliable ``black box''.

The appendices cover various small improvements in the original
algorithms given in {\it AG} and compare the algorithms presented here
with recent work by \citet{nowak22}.

\section{Geodesics in terms of elliptic integrals} \label{elliptic-sec}

\subsection{Formulation of Legendre and Cayley}

One of the earliest systematic studies of geodesics on an ellipsoid was
given by \citet[\S\S126--129]{legendre11} who cast the solution in terms
of his elliptic integrals.  However, when Bessel later wished to obtain
numerical results for the course of a geodesic, he stated ``Because the
tools to compute these special functions [elliptic integrals] are not
yet sufficiently versatile, we instead develop a series solution which
converges rapidly because $e^2$ is so small.''  Bessel's approach,
series expansions valid for small flattening, became the mainstay for
solving the geodesic problem.  For a long time, elliptic integrals
continued to be difficult to compute conveniently; for
example, \citet[Chap.~19]{abramowitz64} only provides tables for the
elliptic integrals for selected values of the modulus and parameter.
This obstacle was only removed with the work of \citet{carlson95} as we
describe below.

Before we give Legendre's solution, let us review the formulation in
terms of the {\it auxiliary sphere}; see Fig.~2 of {\it AG}.  This is a
construct used implicitly by Legendre and explicitly by Bessel that maps
the path of the geodesic on the ellipsoid onto a great circle on the
sphere.  The latitude $\phi$ is mapped to the parametric latitude
$\beta$, the azimuth $\alpha$ is preserved, and $\lambda$ and $s$ are
related to the corresponding spherical variables $\omega$ and $\sigma$
by \citep[Appendix A]{karney11a}
\begin{equation}\label{s-lambda-diff}
\frac1a \frac{\d s}{\d\sigma} = \frac{\d\lambda}{\d\omega} =
\sqrt{1-e^2\cos^2\beta}.
\end{equation}
Substituting the results from spherical trigonometry,
\begin{align}
\frac{\d\omega}{\d\sigma} &= \frac{\sin\alpha}{\cos\beta}
= \frac{\sin\alpha_0}{\cos^2\beta}, \displaybreak[0]\\
\sin\beta &= \cos\alpha_0\sin\sigma,
\end{align}
we obtain
\begin{align}
\label{s-int}
\frac sb &= \int_0^\sigma \sqrt{1 + k^2\sin^2\sigma'}\,\d\sigma',
\displaybreak[0]\\
\label{lambda-int}
\lambda &= (1-f)\sin\alpha_0 \int_0^\sigma
\frac{\sqrt{1 + k^2\sin^2\sigma'}}{1-\cos^2\alpha_0\sin^2\sigma'}\,\d\sigma',
\displaybreak[0]\\
J(\sigma) &= \int_0^\sigma \frac{k^2\sin^2\sigma'}{\sqrt{1 + k^2\sin^2\sigma'}}
\,\d\sigma',
\end{align}
where $k = e' \cos\alpha_0$.  $J(\sigma)$ is a term appearing in the
expressions for the reduced length $m$ and the geodesic scale $M$; both
these quantities, which were introduced by \citet{gauss27-en}, describe
the behavior of nearby geodesics; see Sec.~3 of {\it AG}.  These
expressions can be put in the form of Legendre's elliptic integrals,
\begin{align}
\label{s-ell}
\frac sb &= E(\sigma, ik), \displaybreak[0]\\
\label{lambda-ell}
\lambda &= (1 - f) \sin\alpha_0  G(\sigma, \cos^2\alpha_0, ik),
\displaybreak[0]\\
J(\sigma) &= E(\sigma, ik) - F(\sigma, ik),
\end{align}
where
\begin{align}
G(\phi,\alpha^2,k) &= \int_0^\phi
 \frac{\sqrt{1 - k^2\sin^2\theta}}{1 - \alpha^2\sin^2\theta}\,\d\theta
\notag\\
 &=\frac{k^2}{\alpha^2}F(\phi, k)
 +\biggl(1-\frac{k^2}{\alpha^2}\biggr)\Pi(\phi, \alpha^2, k),
\end{align}
and $F(\phi, k)$, $E(\phi, k)$, and $\Pi(\phi, \alpha^2, k)$, are
incomplete elliptic
integrals \citep[\S\dlmf{19.2(ii)}{19.2.ii}]{dlmf10}.
Equations~(\ref{s-ell}) and (\ref{lambda-ell}) are similar to the
expressions given by Legendre; the only significant difference is that
our formulas involve an imaginary modulus, $ik$, because we choose our
origin for geodesics, $\sigma = 0$, to be at the node (rather than at
the vertex which was Legendre's choice).  This is inconsequential
because the definitions of the elliptic integrals involve the square of
the modulus; furthermore, for {\it prolate} ellipsoids, $ik$ is real.

Equations (\ref{lambda-int}) and (\ref{lambda-ell}) for $\lambda$ are
inconvenient to use in practice because the integrands are arbitrarily
large as the geodesic passes close to a pole leading to a nearly
discontinuous change in $\lambda$.  Of course, this is the behavior of
great circles as they approach a pole, and it is easily described by
spherical trigonometry.  It is therefore advantageous to split the
expression for $\lambda$ into two terms, one involving $\omega$ to
describe the basic behavior plus another, proportional to $f$ and
involving special functions, to describe the ellipsoidal correction.
This can be achieved by integrating
\begin{equation}
\frac{\d(\lambda-\omega)}{\d\omega} = \sqrt{1-e^2\cos^2\beta} - 1,
\end{equation}
which gives
\begin{equation}\label{lambda-int-a}
\lambda=\omega - e^2\sin\alpha_0
\int_0^\sigma \frac1{1+(1-f)\sqrt{1 + k^2\sin^2\sigma'}}
\,\d\sigma',
\end{equation}
where the near-singular behavior of $\lambda$ is captured by $\omega$
with the difference given by a well behaved integral.  This is the
starting point for Bessel's series solution.

\citet{cayley70} found a similar way to transform
Eq.~(\ref{lambda-ell}) to give
\begin{equation}\label{lambda-ell2}
\lambda = \chi
- \frac{e'^2}{\sqrt{1+e'^2}}\sin\alpha_0 H(\sigma, -e'^2, ik),
\end{equation}
where
\begin{align}
\tan\chi &= \sqrt{\frac{1+e'^2}{1+k^2\sin^2\sigma}}\tan\omega,
\displaybreak[0]\\
H(\phi, \alpha^2, k)
&= \int_0^\phi
   \frac{\cos^2\theta}{(1-\alpha^2\sin^2\theta)\sqrt{1-k^2\sin^2\theta}}
   \,\d\theta \notag\\
&=
\frac1{\alpha^2} F(\phi, k) +
\biggl(1 - \frac1{\alpha^2}\biggr) \Pi(\phi, \alpha^2, k).
\end{align}
Now the main contribution to $\lambda$ is $\chi$, a variable that
behaves similarly to the longitude near a pole, and the elliptic
integral is relegated to a correction term; not only is it multiplied by
$e'^2$ but the parameter of the elliptic integral, $-e'^2$ is also
small.  This allows the longitude to be computed more accurately.  The
equivalence of the two expressions for $\lambda$,
Eqs.~(\ref{lambda-ell}) and (\ref{lambda-ell2}), follows
from \citet[Eq.~\dlmf{19.7.8}{19.7.E8}]{dlmf10}.

\subsection{Numerical evaluation of the elliptic integrals}\label{num-ell}

\citet{carlson95} provided high quality algorithms that allowed
symmetric elliptic integrals \citep[\S\dlmf{19.16}{19.16}]{dlmf10} to be
computed to arbitrary accuracy.  Legendre's elliptic integrals can then
be simply evaluated using \citet[\S\dlmf{19.25(i)}{19.25.i}]{dlmf10}.
In our implementation, we also use the amendments to Carlson's paper
given in \citet[\S\dlmf{19.36(i)}{19.36.i}]{dlmf10}.

With these methods in hand, we review the solutions of the direct and
inverse geodetic problems.  We first discuss the direct problem,
determining the final position $\phi_2$, $\lambda_2$ and azimuth
$\alpha_2$, given a starting position $\phi_1$, $\lambda_1$, initial
azimuth $\alpha_1$, and distance $s_{12}$.  The starting information
allows us to calculate the parameter characterizing the geodesic,
$\alpha_0$.  Equations (\ref{s-ell}) and (\ref{lambda-ell2}) above
provide a reliable way to follow a geodesic arbitrarily far given its
starting position and initial azimuth.  The independent parameter in
these equations is $\sigma$.  Because we are given the distance $s_{12}$
to the second point, we need to invert Eq.~(\ref{s-ell}) to find
$\sigma$ in terms of $s$.  This presents no difficulty; $s$ is a
monotonically increasing function of $\sigma$ so there's a unique
solution and, provided we're sufficiently close to the solution,
Newton's method can be used to obtain an accurate result; the derivative
needed for Newton's method, $\d s/\d\sigma$ is given by
Eq.~(\ref{s-lambda-diff}).  This allows the direct geodesic problem to
be solved in the same way as in {\it AG}.

\begin{figure}[tb]
\begin{center}% produced by geod_nonstand.m
\includegraphics[scale=0.75]{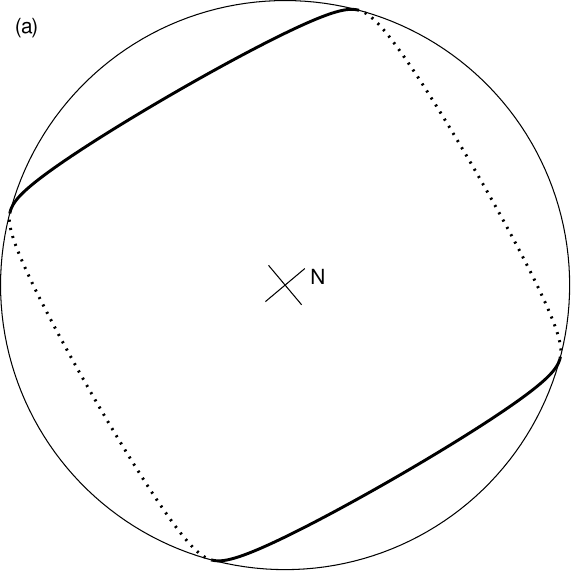}\\[5ex]
\includegraphics[scale=0.75]{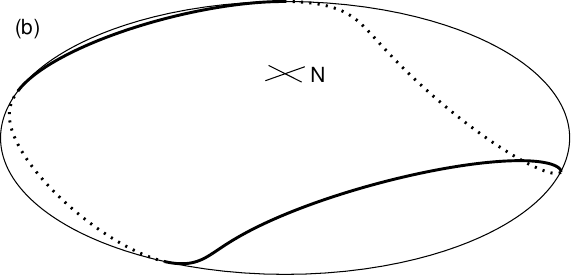}
\end{center}
\caption{\label{geod-nonstand}
A simple closed geodesic for an ellipsoid with $b/a=\frac14$.  The
geodesic starts on the equator with azimuth $\alpha_0 \approx
51.24052^\circ$ and completes 2 complete undulations about the equator
before returning to the starting longitude: (a)~a top view; (b)~a view
at an inclination of $25^\circ$ to the equatorial plane; ``N'' marks the
north pole.  The solid and dotted lines show the visible and hidden
portions of the geodesic.}
\end{figure}%
Figure \ref{geod-nonstand} illustrates a simple (i.e.,
non-self-intersecting) closed geodesic on an ellipsoid with $f
= \frac34$ or $b/a=\frac14$.  This is one of a class of ``non-standard''
closed geodesics that are neither equatorial nor meridional and which
appear for sufficiently oblate ellipsoids, $a >
2b$ \citep[\S3.5.19]{klingenberg82}.

\subsection{Solving the inverse problem}

The inverse problem is more challenging.  Here we are given the
endpoints $\phi_1$, $\lambda_1$ and $\phi_2$, $\lambda_2$, and wish to
determine the shortest distance $s_{12}$ and azimuths $\alpha_1$ and
$\alpha_2$.  Because we have no direct way to determine $\alpha_0$, we
must resort to an iterative solution.  The traditional method for
solving this problem, given by \citet{helmert80-en} and used
by \citet{vincenty75a}, is functional iteration which fails to converge
in cases where the endpoints are nearly antipodal.

A key contribution of {\it AG} was to reduce the determination of
$\alpha_0$ to a simple one-dimensional root-finding problem.  I briefly
recapitulate the method here.  Let's assume we have already dealt with
equatorial and meridional geodesics.  Equatorial geodesics are those
with $\phi_1 = \phi_2 = 0$ and, for oblate ellipsoids, the additional
condition that $\abs{\lambda_{12}} \le (1-f)\pi$; for these we
have $\alpha_1 = \pm\frac12\pi$.  Meridional geodesics are those with
$\lambda_{12} = 0$ or $\pm\pi$ and, for prolate ellipsoids, the
additional condition that $m_{12} \ge 0$; and for these geodesics, we
have $\alpha_1 = 0$ or $\pm\pi$.

For the remaining cases, we can, without loss of generality, specialize
to the case
\begin{equation}
\phi_1 \le 0,\quad \phi_1 \le \phi_2 \le -\phi_1,\quad
 0 < \lambda_{12} \le \pi.
\end{equation}
Note that $\lambda_{12} = 0$ is excluded---this case corresponds to a
meridional geodesic; the case $\lambda_{12} = \pi$ only arises for $f <
0$ (for $f \ge 0$ this also corresponds to a meridional geodesic).

For a particular trial azimuth $\alpha^*_1$,
we solve the {\it hybrid} problem: given
$\phi_1$, $\alpha^*_1$, and $\phi_2$, compute the longitude difference
$\lambda_{12}(\alpha^*_1)$ corresponding to the first intersection of the
geodesic with the circle of latitude $\phi_2$.  The solution of the
inverse problem is then given by finding $\alpha^*_1 = \alpha_1$ which solves
\begin{equation}\label{hybrid-eq}
\lambda_{12}(\alpha_1) = \lambda_{12},
\end{equation}
with $0 < \alpha_1 < \pi$.  We use $\alpha_1$ as the control variable
for this root-finding problem; this is a more convenient choice than
$\alpha_0$ and, once this is found, we can readily obtain $\alpha_0$.
In {\it AG}, we solved this equation using Newton's method.  This
converges very rapidly but requires a good starting guess for the
solution and this, in turn, required a careful analysis of the case of
nearly antipodal endpoints.

\begin{figure}[tb]
\begin{center}% produced by hybrid.m
\includegraphics[scale=0.75]{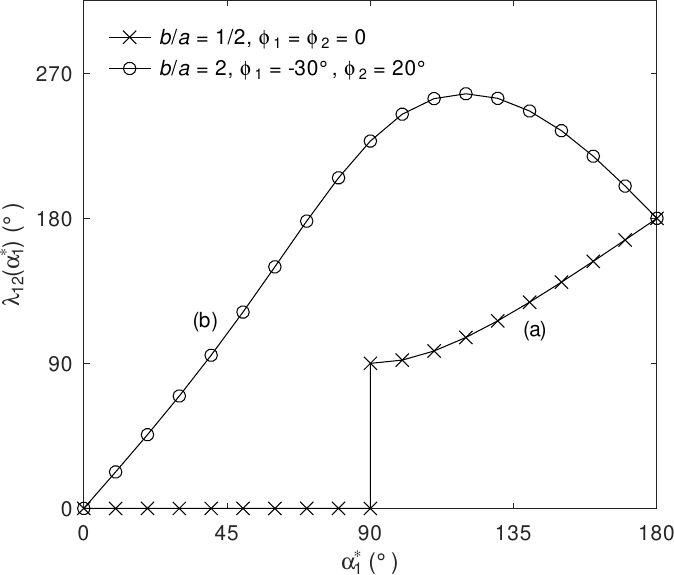}
\end{center}
\caption{\label{hybrid-fig}
The solution to the hybrid problem for (a)~$\phi_1 = \phi_2 = 0$ and
$b/a = \half$ (marked with crosses); (b)~$\phi_1 = -30^\circ$, $\phi_2 =
20^\circ$, and $b/a = 2$ (marked with circles).  This illustrates the
functional dependence of $\lambda_{12}(\alpha^*_1)$.  }
\end{figure}%
This technique worked well with ellipsoids with small flattening.
However, in the general case, the starting guess was sometimes
sufficiently far from the true solution that Newton's method failed.
Before describing how we might deal with this, let us first review the
properties of $\lambda_{12}(\alpha^*_1)$.  First of all, we have
$\lambda_{12}(0) = 0$ and $\lambda_{12}(\pi) = \pi$.  Furthermore, with
one exception, $\lambda_{12}(\alpha^*_1)$ varies continuously so
Eq.~\ref{hybrid-eq} is guaranteed to have a solution.  The exception is
illustrated in Fig.~\ref{hybrid-fig}(a); this is the case with endpoints
on the equator for an oblate ellipsoid and $\lambda_{12}(\alpha^*_1)$ is
discontinuous at $\alpha^*_1 = \half\pi$.  Because equatorial and
meridional geodesics have already been handled, this only occurs for
${(1-f)\pi} < \lambda_{12} < \pi$ and we can restrict the possible range
of azimuth to $\half\pi < \alpha^*_1 < \pi$.  Within this range,
$\lambda_{12}(\alpha^*_1)$ is continuous and a solution exists.  Finally,
there is exactly {\it one} root and hence the derivative
$\d\lambda_{12}(\alpha^*_1)/\d\alpha^*_1 = \lambda_{12}'(\alpha^*_1)$ is
non-negative at that root.  In this context, it is important to consider
the case of a prolate ellipsoid; see Fig.~\ref{hybrid-fig}(b).  Even
though $\lambda_{12}(\alpha^*_1)$ has an interior maximum,
$\lambda_{12}(\alpha^*_1) = \lambda_{12}$ still has only a single root
provided that $\lambda_{12} \le \pi$ (stipulated in the initial setup
of the inverse problem) and that $\alpha_1 < \pi$ (because we have
already handled meridional geodesics).

With this background, even the crudest root-finding techniques, e.g.,
bisection or {\it regula falsi}, can be used to find the solution.  We
shall continue to use Newton's method on account of its rapid
convergence.  However we back it up with the bisection method to deal
with two possible problems with Newton's method:
(a)~$\lambda_{12}'(\alpha^*_1)$ is so small that the new value of $\alpha^*_1$
is outside the range $(0,\pi)$ or (b)~$\lambda_{12}'(\alpha^*_1)$ is
negative, in which case the new value of $\alpha^*_1$ is further from the
true root.  During the course of the Newton iteration, we maintain a
bracketing range for $\alpha_1 \in (\alpha_{1-}, \alpha_{1+})$.
Initially, we have $\alpha_{1-} = 0$, $\alpha_{1+} = \pi$.  On each
iteration, we update one of $\alpha_{1\pm}$ to $\alpha^*_1$ depending on
the sign of $\lambda_{12}(\alpha^*_1) - \lambda_{12}$, provided that this
results in a tighter bracket for $\alpha_1$.  Whenever either of the
problems, (a) or (b) listed above, occurs, we restart Newton's method
with a new starting guess $\alpha^*_1 = \half(\alpha_{1-} + \alpha_{1+})$;
and we will be able to bisect the bracket after the next iteration.

We can use the spherical solution to get a starting value for $\alpha^*_1$
with a value of $0$ or $\pi$ (i.e., outside the allowed range for
$\alpha_1$) replaced by $\frac12\pi$.  (See also
Appendix~\ref{equatorial-ends} for additional caveats for the case of
equatorial endpoints.)  Typically, the bisection safety mechanism is
invoked at most once when solving the inverse problem.  The net result
is that this provides a reliable (always converging) and fast (rapidly
converging) method of solving the inverse geodesic problem for arbitrary
ellipsoids.  In some cases, the inverse problem does not have a unique
solution---there are multiple shortest geodesics; these cases are
readily handled as discussed in Appendix~\ref{multiple-shortest}.

\section{The area integral}\label{area-sec}

\subsection{Formulation of Danielsen}

In {\it AG}, $S_{12}$ was defined as the area between a geodesic segment
$s_{12}$ and the equator, i.e., the area of the geodesic quadrilateral
$AF\!H\!B$ in Fig.~1 of {\it AG}.  Following \citet{danielsen89}, this
was cast as an integral as follows:
\begin{equation}\label{Sdef}
S_{12} = S(\sigma_2) - S(\sigma_1),\quad
S(\sigma) = c^2\bigl(\alpha + p(\sigma)\bigr),
\end{equation}
where
\begin{equation}
c^2 =
\frac{a^2}2 + \frac{b^2}2 \frac{\tanh^{-1}e}e
\end{equation}
is the authalic radius squared, and
\begin{align}
\label{i4int}
p(\sigma) &= \int_{\pi/2}^\sigma q(\sigma')\,\d\sigma',\displaybreak[0]\\
q(\sigma) &= - A_4 \Delta t(e'^2, k^2\sin^2\sigma) \frac{\sin\sigma}2,
\displaybreak[0]\\
A_4 &= \frac{e^2a^2}{c^2}\cos\alpha_0 \sin\alpha_0,\label{a4-def}
\displaybreak[0]\\
\Delta t(x,y) &= \frac{t(x)-t(y)}{x-y},\displaybreak[0]\\
t(x) &= x +  \sqrt{1+x}\, \frac{\sinh^{-1}\sqrt x}{\sqrt x}.
\end{align}
This is a change from the notation in Eq.~(59) of {\it AG}; the
relationship is
\begin{equation}
p(\sigma) = A_4 I_4(\sigma);
\end{equation}
this simplifies the assessment of the error in $S_{12}$.  The expression
$\Delta t(x,y)$ is the {\it divided difference} of $t(x)$.  In the limit
$y \rightarrow x$, we have $\Delta t(x,x)=t'(x)$.  If $x$ and $y$ are
distinct and have the same sign, there's a systematic way to express
$\Delta t(x,y)$ in a form that avoids excessive roundoff
errors \citep{kahan99}.

The area of an arbitrary geodesic polygon is easily computed by summing
the signed area contributions $S_{12}$ for each edge.  An adjustment to this
total needs to be made if the polygon encircles a pole as described in
{\it AG}.

The challenge is that the integral, Eq.~(\ref{i4int}), cannot be carried
out in closed form.  However both $q(\sigma)$ and $p(\sigma)$ are
periodic functions and so may be written as Fourier series:
\begin{align}
q(\sigma) &= \sum_{l=0}^\infty Q_l \sin\bigl(2(l+\half)\sigma\bigr),
\label{fourier-series}
\displaybreak[0]\\
\label{fourier-int}
Q_l &= \frac4\pi \int_0^{\pi/2} q(\sigma)
\sin\bigl(2(l+\half)\sigma\bigr)\,\d\sigma,\displaybreak[0]\\
p(\sigma)&= \sum_{l=0}^\infty P_l \cos\bigl(2(l+\half)\sigma\bigr),
\label{p-series}\displaybreak[0]\\
P_l &= -\frac{Q_l}{2(l+\frac12)}.\label{c4-eq}
\end{align}
Here, Eqs.~(\ref{fourier-series}) and (\ref{fourier-int}) constitute a
Fourier transform pair.  In the next sections, we explore two ways to
compute $P_l$ approximately.

\subsection{Using a Taylor series}\label{taylor-sec}

In {\it AG}, we gave Taylor series expansions for $P_l$ using $e'^2$ and
$k^2$ as small parameters.  For reasons explained in
Appendix~\ref{areaseries}, this is a poor choice of expansion parameters
for eccentric ellipsoids; so we replace the expansion parameters by $n$
and $\epsilon$, Eqs.~(\ref{eps-def}), resulting in the expansion
Eqs.~(\ref{c4-series}) which should be used instead of {\it AG}'s
Eqs.~(63).  This expansion may be used to approximate $p(\sigma)$ with
\begin{align}
\tilde p^{(N)}(\sigma)&=
\sum_{l=0}^\infty \tilde P_l^{(N)} \cos\bigl(2(l+\half)\sigma\bigr),
\displaybreak[0]\\
\tilde P_l^{(N)} &= A_4 C_{4l},
\end{align}
where $A_4$ is defined in Eq.~(\ref{a4-def}) and the superscript $(N)$
gives the order at which the Taylor series is truncated; thus
Eqs.~(\ref{c4-series}) with the ellipses dropped corresponds to $N=6$.
$A_4$ is proportional to $e^2$, so the highest order terms in
Eqs.~(\ref{c4-series}) are $O(f^6)$; furthermore, we have $\tilde
P_l^{(N)} = 0$ for $l\ge N$.

One possibility for computing the area on eccentric ellipsoids is
extending the Taylor series expansion to higher order.  Thus, when I
first implemented the solution of the geodesic problem in terms of
elliptic integrals as described in Sec.~\ref{elliptic-sec}, I computed
the area using $\tilde p^{(30)}(\sigma)$; this gives full double
precision accuracy for $\half \le b/a \le 2$.  However, this approach
quickly becomes cumbersome for extremely eccentric ellipsoids:
\begin{itemize}
\ifeprint\itemsep-0.4ex plus 0.5ex \fi %%
\item the value of $N$ required to give full accuracy becomes very large;
\item the cost of determining the
coefficients in the Taylor series becomes impractical---it
takes \citet{maxima} 17 minutes for $N=32$ (tolerable for generating
code) but 19 hours to extend this to $N=64$ (which is starting to get
painful);
\item the storage required for the coefficients becomes excessive,
scaling as $N^3/6$;
\item once the ellipsoid is selected,
evaluating the coefficients of $\epsilon^j$ in Eqs.~(\ref{c4-series})
requires $O(N^3)$ operations;
\item once $\alpha_0$ is given,
evaluating $\tilde P_l^{(N)}$ requires $O(N^2)$ operations.
\end{itemize}

\subsection{Quadrature using discrete sine transforms}\label{quad-dst}

To overcome the shortcomings of Taylor series for large eccentricities,
I explored evaluating Eq.~(\ref{i4int}) using the quadrature method
of \citet{clenshaw60}.  Given a definite integral,
\[
\int_{-1}^1 g(x)\, \d x,
\]
this method entails expanding $g(x)$ as Chebyshev polynomials, or,
equivalently, $g(\cos\theta)$ as a Fourier series.  This allows the
definite integral to be computed to high accuracy and, as a bonus, the
indefinite integral can be computed.  The accuracy is a result of the
fast convergence of the trapezoidal rule for integrals of a periodic
function over a full period; in this case, the integrals give the
Fourier coefficients.

However, this is a roundabout procedure given that we are starting with
a periodic function.  Instead, we directly evaluate
Eq.~(\ref{fourier-int}) using the trapezoidal rule and evaluate
$p(\sigma)$ by using the resulting Fourier coefficients in
Eqs.~(\ref{p-series}) and (\ref{c4-eq}).  Thus we approximate $Q_l$ with
\begin{equation}\label{dst-iii}
\hat Q_l^{(N)} = \frac{4h_N}\pi \sump_{j=0}^{N-1} q((j+1)h_N)
\sin\bigl(2(l+\half)(j+1)h_N\bigr),
\end{equation}
where $h_N = \pi/(2N)$, we take $0 \le l < N$, and the prime on the
summation sign indicates that the last term in the sum should be halved.
Because $q(\sigma)$ is an analytic periodic function, $\hat Q_l^{(N)}$
converges to the true value $Q_l$ faster than any power of $N$, as
$N\rightarrow\infty$.  Equation (\ref{dst-iii}) is, in fact, the
discrete sine transform, DST, of type III \citep{wang85}.  Its inverse
is a type II DST,
\begin{equation}
q((j+1)h_N) = \sum_{l=0}^{N-1} \hat Q_l^{(N)}
\sin\bigl(2(l+\half)(j+1)h_N\bigr);
\end{equation}
this holds for all integer $j$.  Rewriting this as a continuous function,
\begin{equation}
\hat q^{(N)}(\sigma) = \sum_{l=0}^{N-1} \hat Q_l^{(N)}
\sin\bigl(2(l+\half)\sigma\bigr),
\end{equation}
we see that $\hat q^{(N)}(\sigma)$ is likely to be an excellent
approximation to $q(\sigma)$ given that the functions coincide wherever
$\sigma$ is a multiple of $h_N$.  In the same way, we approximate
$p(\sigma)$ by
\begin{align}\label{p-sum}
\hat p^{(N)}(\sigma)&=
\sum_{l=0}^\infty \hat P_l^{(N)} \cos\bigl(2(l+\half)\sigma\bigr),
\displaybreak[0]\\
\hat P_l^{(N)} &= \begin{cases}
\displaystyle
-\frac{\hat Q_l^{(N)}}{2(l+\frac12)},&\text{for $l < N$},\\
0, &\text{for $l \ge N$}.
\end{cases}
\end{align}

A convenient way to compute DSTs is using the fast Fourier transform,
FFT \citep{cooley65}.  This reduces the cost of evaluating
Eq.~(\ref{dst-iii}) for $0\le l <N$ from $O(N^2)$ to $O(N\log N)$.
Furthermore, \citet{gentleman72b} points out that the FFT results in
smaller roundoff errors compared with other methods.  Equation
(\ref{p-sum}) can be evaluated using \citet{clenshaw55} summation; this
method is fast (the cosine terms are computed by a recurrence relation)
and because $\hat Q_l^{(N)}$ decays rapidly with $l$ and because the
summation starts at high values of $l$, it is also accurate.

As we shall see, the error $\max(\abs{ \hat q^{(N)}(\sigma)
-q(\sigma) })$ decays exponentially with $N$; thus doubling $N$
roughly squares the error.  Furthermore, it's possible to double $N$
rather inexpensively.  First, generate the type IV DST for $q(\sigma)$,
\begin{equation}
\check Q_l^{(N)} = \frac{4h_N}\pi \sum_{j=0}^{N-1} q((j+\half)h_N)
\sin\bigl(2(l+\half)(j+\half)h_N\bigr).
\end{equation}
The sums for computing $\hat Q_l^{(N)}$, Eq.~(\ref{dst-iii}), and
$\check Q_l^{(N)}$ involve evaluating $q(\sigma)$ at intervals of $\half
h_N = h_{2N}$; it follows that $\hat Q_l^{(2N)}$ is just the mean of
$\hat Q_l^{(N)}$ and $\check Q_l^{(N)}$.  These coefficients for $N\le l
< 2N$ are
\begin{equation}
\hat Q_l^{(N)} = - \hat Q_{2N-1-l}^{(N)},\quad
\check Q_l^{(N)} = + \check Q_{2N-1-l}^{(N)}.
\end{equation}
Thus, $\hat Q_l^{(2N)}$ is given by
\begin{equation}\label{2n-eq}
\hat Q_{N-1/2\pm g}^{(2N)} =
\frac{\check Q_{N-1/2-\abs g}^{(N)} \mp \hat Q_{N-1/2-\abs g}^{(N)}}2,
\end{equation}
for $g = \frac12, \frac32, \ldots, N - \frac12$.  This method avoids
wasting the results $\hat Q_l^{(N)}$ which have already been
computed \citep{gentleman72b}.

\subsection{Convergence of approximate Fourier series}

\begin{figure}[tb]
\begin{center}% produced by c4conv.m
\includegraphics[scale=0.75]{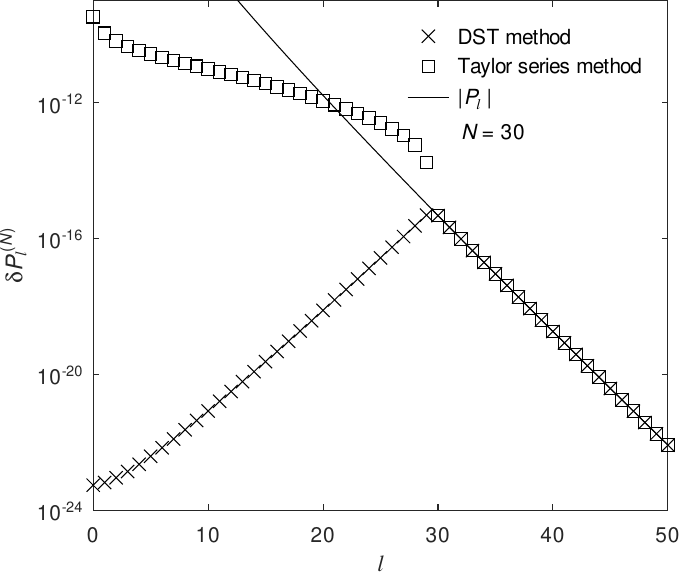}
\end{center}%
\caption{\label{deltaP-l}
The dependence of the errors in the individual Fourier coefficients,
$P_l$, on $l$ for $N=30$.  The crosses, resp.~squares, show the errors
using the DST method, $\delta\hat P_l^{(N)}$, resp.~the Taylor series
method, $\delta\tilde P_l^{(N)}$.  The case illustrated is for a geodesic
on an ellipsoid with $b/a=\frac14$, $\alpha_0=\frac14\pi$.  The line
shows the exact value of $\abs{P_l}$.}
\end{figure}%
In assessing how well $\hat p^{(N)}(\sigma)$, Eq.~(\ref{p-sum}),
approximates $p(\sigma)$, we start by looking at the error in the
individual Fourier coefficients, namely $\delta \hat P_l^{(N)}
= \abs{ \hat P_l^{(N)} - P_l }$ for a given $N$.  For illustrative
purposes, we consider $b/a = \frac14$, $\alpha_0 = \frac14 \pi$, and
consider $N = 30$; we compute the ``exact'' $P_l$ using a much larger
value $N$.  The resulting errors $\delta \hat P_l^{(30)}$ are shown in
Fig.~\ref{deltaP-l} as crosses.  The figure also shows $\abs{P_l}$ (the
solid line) decaying exponentially as a function of $l$.  Obviously
$\delta \hat P_l^{(30)}$, which equals $\abs{P_l}$ for $l\ge
30$, shows the same exponential decay.  However, we also have
exponential decay in $\delta \hat P_l^{(30)}$ as $l$ {\it decreases}
from $29$ to $0$; indeed, $\delta \hat P_l^{(30)}$ is roughly symmetric
about $l = 30$.  This behavior is predicted by Eq.~(\ref{2n-eq}), since
we can approximate $\delta \hat P_l^{(N)}$ by $\abs{\hat P_l^{(N)}
- \hat P_l^{(2N)}}$, with
\begin{equation}
\hat P_{N-1/2\pm g}^{(N)} - \hat P_{N-1/2\pm g}^{(2N)} =
\frac{\hat Q_{N-1/2-\abs g}^{(N)}
  - \check Q_{N-1/2-\abs g}^{(N)}}{4(N\pm g)};
\end{equation}
the variation of the numerator in this expression (symmetric in $g$) is
much stronger than the variation of the denominator.  We conclude that
the error arising from using approximate Fourier coefficients, $\hat
P_l^{(N)}$ instead of $P_l$, is approximately the same as the error from
truncating the Fourier sum in Eq.~(\ref{p-sum}).

Figure~\ref{deltaP-l} also shows the errors in the Taylor series
coefficients $\delta \check P_l^{(N)} = \abs{\check P_l^{(N)} - P_l}$
for $N=30$ as squares.  The error is, of course, the same as for
$\delta \hat P_l^{(30)}$ for $l \ge 30$; but for $l < 30$, the error is
several orders of magnitude bigger.

\begin{figure}[tb]
\begin{center}% produced by c4conv.m
\includegraphics[scale=0.75]{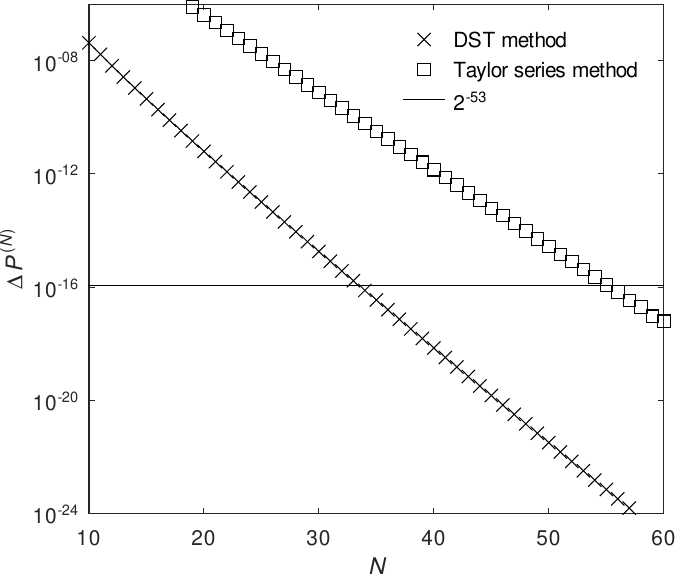}
\end{center}
\caption{\label{DeltaP-N}
The dependence of the overall error in the Fourier series approximation
to $p(\sigma)$ on $N$.  The crosses, resp.~squares, show the errors for
the DST method, $\Delta \hat P^{(N)}$, resp.~the Taylor series method,
$\Delta \tilde P^{(N)}$.  This is for the same case as
Fig.~\ref{deltaP-l}, namely $b/a=\frac14$, $\alpha_0=\frac14\pi$.  The
horizontal line indicates the roundoff limit for double precision
$2^{-53} \approx 10^{-16}$.}
\end{figure}%
Figure \ref{deltaP-l} gives the errors in the individual Fourier
coefficients for a particular $N$.  To gauge how the overall error in
$\hat p^{(N)}(\sigma)$, Eq.~(\ref{p-sum}), varies with $N$, we combine
the errors in the coefficients, as follows,
\begin{equation}
\Delta \hat P^{(N)} = \sum_{l=0}^\infty \delta \hat P_l^{(N)}.
\end{equation}
We similarly define $\Delta \tilde P^{(N)}$ for the
coefficients obtained by Taylor expansion.  Figure \ref{DeltaP-N} shows
how these error metrics vary as $N$ is increased from $10$ to $60$ for
$b/a = \frac14$ and $\alpha_0 = \frac14\pi$ (the same parameters used
for Fig.~\ref{deltaP-l}).  Both show exponential decay with $\Delta \hat
P^{(N)}$ being several orders of magnitude smaller and decaying at a
slightly faster rate.

\subsection{Determining the number of points in the quadrature}

In practical applications, we will evaluate the area using
Eqs.~(\ref{Sdef}) and have to contend both with truncation errors,
because we use a finite value of $N$ in computing $p(\sigma)$, and with
roundoff errors, because of the limited precision of the floating-point
number system.  We would like to pick $N$ large enough that the
truncation errors are less than the roundoff errors and for this we
require that $\Delta \hat P^{(N)} \le 2^{-m}$ where $m$ is the number of
bits in the fraction of a floating-point number.  For double precision,
we typically have $m = 53$ and, substituting the value of $c\approx
6371\,\mathrm{km}^2$ for the earth, this gives a truncation error in
$S(\sigma)$ of $2^{-53}c^2 \approx 50\,\mathrm{cm}^2$.  The horizontal
line in Fig.~\ref{DeltaP-N} labels this error threshold, $2^{-53}$.
This shows that, for this case, we need $N\ge 34$ for the DST method
compared to $N\ge 56$ when using a Taylor expansion.

\begin{figure}[tb]
\begin{center}% produced by ndouble.m
\includegraphics[scale=0.75]{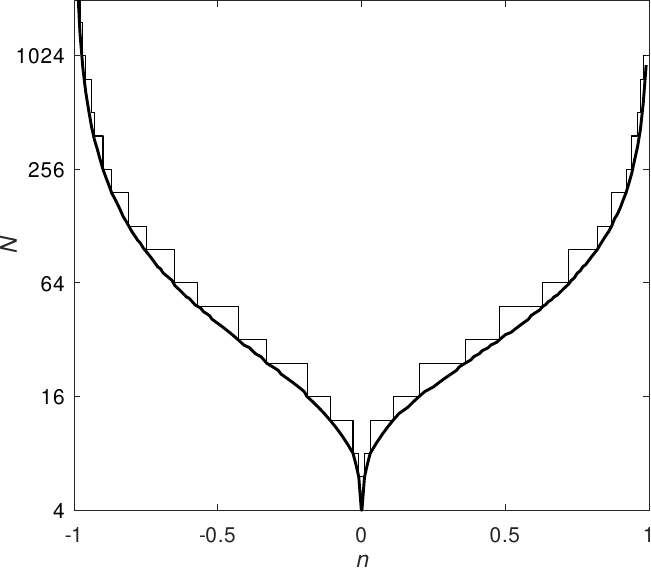}
\end{center}
\caption{\label{N-double}
The heavy line shows the minimum number of points $N$ required in the
DST to ensure double precision accuracy as a function of $n$ for $\abs
n \le 0.99$ and for arbitrary $\alpha_0$.  If we limit the
possible values of $N$ to $2\times 2^j$ and $3\times 2^j$ with integer
$j \ge 1$, then the result is the staircase shown as a light line.}
\end{figure}%
In principle, we could reestimate $N$ for each geodesic, i.e., for every
$n$ and $\alpha_0$.  (Since we are dealing with both oblate and prolate
ellipsoids of arbitrary eccentricity, we use $n$ as the measure of
eccentricity instead of $e = 2\sqrt n/(1+n)$ or $f = 2n/(1+n)$.)
However, that would require extra machinery to do the estimation at run
time.  Instead, we estimate $N$ for $-0.99 \le n \le 0.99$.  For a
particular $n$, we find the minimum $N$ that satisfies the accuracy
requirement for {\it all} $\alpha_0$.  The result is shown in
Fig.~\ref{N-double} as the heavy line.  The light line shows the result
of restricting the allowed values of $N$ to $2\times 2^j$ and $3\times
2^j$ with integer $j \ge 1$; now $N$ is the product of small factors
($2$ and $3$), a requirement for the efficient implementation of the
FFT.  The light line is converted to a simple table allowing $N$ to be
looked up at run time based on the value of $n$.

The conclusion is that the discrete sine transform method provides a
convenient and accurate way to evaluate the area integral.  In
comparison to the Taylor series method: the same accuracy is obtained
with substantially fewer terms ($N = 34$ for the DST versus $N = 56$ for
the Taylor series in the example illustrated in Fig.~\ref{DeltaP-N}).
Just as important, $N$ can be easily adjusted for a particular $n$,
Fig.~\ref{N-double} (no need to compute a new Taylor series), the memory
required is $O(N)$ instead of $O(N^3)$, and the setup time for a
particular geodesic is $O(N\log N)$ instead of $O(N^3)$.

\section{Implementation}\label{implement}

The algorithms detailed in Secs.~\ref{elliptic-sec} and \ref{area-sec}
have been implemented in version 2.1.2 of the open-source C++ library,
GeographicLib \citep{geographiclib212}.  The solution of the geodesic
problems in terms of elliptic integrals was added in version 1.25
(2012).  At that time, the area was computed with a 30th order Taylor
series as described in Sec.~\ref{taylor-sec}.  This was replaced by the
implementation using DSTs, Sec.~\ref{quad-dst}, in version 2.1 (2022).
These algorithms {\it supplemented} the earlier ones described in {\it
AG} since, as we see below, these remain the preferred methods for
terrestrial ellipsoids.

We use the KISS FFT package \citep{kissfft} for performing the DST.
Even though this package has no direct support for the DSTs, as a
``header-only'' package, it is easier to integrate into GeographicLib
than the more capable FFTW package \citep{fftw}.

The main geodesic calculations are also exposed with two utility
programs provided with GeographicLib, {\tt GeodSolve} for geodesic
calculations, and {\tt Planimeter} for measuring the perimeter and area
of geodesic polygons.  With both utilities, the {\tt -E} flag uses the
algorithms described here for arbitrary ellipsoids.

In developing numerical algorithms, especially those that offer high
accuracy, it is very useful to be able to distinguish truncation errors
from roundoff errors.  For this reason, high precision calculations were
added to GeographicLib in version 1.37 (2014).  This allowed the library
to be compiled with various alternatives to standard double precision
(53 bits of precision or 16 decimal digits) for floating-point numbers:
\begin{itemize}
\ifeprint\itemsep-0.4ex plus 0.5ex \fi %%
\item extended precision (64 bits or 19 decimal digits)
available as the ``long double'' type on some platforms;
\item quadruple precision (113 bits or 34 decimal digits) using the
boost multi-precision library \citep{boost-multiprec};
\item arbitrary precision (selected at the start of execution)
using the MPREAL C++ interface \citep{mpreal} to the MPFR
library \citep{mpfr}.
\end{itemize}
The results in Figs.~\ref{deltaP-l}, \ref{DeltaP-N}, and \ref{N-double}
and in Tables \ref{oblate-tab} and \ref{prolate-tab}, below, were
computed with the last option setting the precision to 256 bits (77
decimal digits).

\section{Results}\label{results}

%% time ./develop/GeodTest -a [or -E] < ../scratch/GeodTest.dat
%% err[0] error in position of point 2 for the direct problem.
%% err[1] error in azimuth at point 2 for the direct problem.
%% err[2] error in m12 for the direct problem & inverse (except near conjugacy)
%% err[3] error in s12 for the inverse problem.
%% err[4] error in the azimuths for the inverse problem scaled by m12.
%% err[5] consistency of the azimuths for the inverse problem.
%% err[6] area error direct & inverse (except near conjugacy)
%%   series dst
%% 0 10.03 28.17
%% 1  9.16 11.67
%% 2 10.72 27.49
%% 3  7.45 22.35
%% 4  3.93  4.73
%% 5  9.54 21.21
%% 6  9.80  9.80
%% time 3.6 8.4
A starting point for testing the new algorithms is the test data set
given in \citet{geodtest}.  This is specific to the WGS84 ellipsoid and
as such allows us to compare the new algorithms with the series
solutions described in {\it AG}.  The maximum errors in geodesic
calculations when converted to a distance are about $0.03\,\mu\mathrm m$
which is 2--3 times larger than the corresponding figures for the series
solution.  The error in the area is about $0.1\,\mathrm m^2$ in both
cases.  On this data set the new routines run about 2.5 times slower.
The lower accuracy and speed are to be expected given the additional
complexity of the new algorithms.

The lesson is clear: for terrestrial ellipsoids, the algorithms described
in {\it AG} are to be preferred---they are faster and more accurate than
the new routines.  In fact, the new routines are {\it still} very
accurate and fast.  Once the eccentricity of the ellipsoid is such that
$\abs f > \frac1{50}$, the accuracy of the series solution degrades and
the new routines should be used.  Let us, therefore, turn to assessing
their performance on eccentric ellipsoids.

\begin{table}[tb]
\caption{\label{oblate-tab}
Results for a geodesic segment with $\phi_1 = 0$, $\lambda_1 = 0$,
$\alpha_1 = \frac14\pi$, $\sigma_{12} = \half\pi$ on an oblate ellipsoid
with $a = 6\,400\,\mathrm{km}$ and with various values of the third
flattening, $n$.  For each value of $n$, the first, resp.~second, row
gives the results of a high precision, resp.~double precision,
calculation.  The data is displayed with 17 decimal digits and all the
digits in the first rows may be regarded as correct.  The differing
digits in the second row are underlined, but differences in the 17th
digit are usually not significant.  The analytic results
for $n = 0$ and $1$ are also given.}
\begin{center}\let\u\underline
\begin{tabular}{>{$\scriptstyle}l<{$}>{$\scriptstyle}r<{$}>
{$\scriptstyle}r<{$}>{$\scriptstyle}r<{$}}
\hline\hline\noalign{\smallskip}
\multicolumn{1}{c}{$n$} & \multicolumn{1}{c}{$\lambda_2\,({}^\circ)$} &
\multicolumn{1}{c}{$s_{12}\,(\mathrm{m})$} &
\multicolumn{1}{c}{$S_{12}\,(\mathrm{m}^2)$}
\\\noalign{\smallskip}\hline\noalign{\smallskip}
\multicolumn{1}{l}{$0$} &
\multicolumn{1}{c}{$90$} &
\multicolumn{1}{c}{$\half\pi a$} &
\multicolumn{1}{c}{$\frac14\pi a^2$} \\
\noalign{\smallskip}
0.01 &88.742968019148302    &9904105.0587012822    &31213542356109.085    \\
     &88.74296801914830\u{7}&9904105.05870128\u{80}&31213542356109.0\u{90}\\
\noalign{\smallskip}
0.02 &87.516869281406891    &9758656.5485393260    &30293469475840.470    \\
     &87.5168692814068\u{83}&9758656.54853932\u{56}&30293469475840.4\u{69}\\
\noalign{\smallskip}
0.05 &84.015774978368889    &9342609.4418442232    &27735488324240.197    \\
     &84.0157749783688\u{96}&9342609.44184422\u{12}&27735488324240.19\u{5}\\
\noalign{\smallskip}
0.10 &78.725380139212172    &8711622.0524734494    &24064301808040.490    \\
     &78.7253801392121\u{63}&8711622.05247344\u{26}&24064301808040.4\u{88}\\
\noalign{\smallskip}
0.20 &69.896175299112817    &7650604.1274847332    &18453877989937.629    \\
     &69.8961752991128\u{26}&7650604.1274847332    &18453877989937.62\u{5}\\
\noalign{\smallskip}
0.40 &57.573823093058582    &6143630.9790943809    &11678681837788.421    \\
     &57.57382309305858\u{6}&6143630.9790943\u{796}&11678681837788.42\u{2}\\
\noalign{\smallskip}
0.60 &50.239779898617183    &5219414.0281749099    &8187126653111.4930    \\
     &50.239779898617\u{222}&5219414.0281749\u{116}&8187126653111.49\u{80}\\
\noalign{\smallskip}
0.90 &45.355849749995502    &4575456.8211684255    &6010343299883.0257    \\
     &45.355849749995\u{450}&4575456.8211684255    &6010343299883.025\u{4}\\
\noalign{\smallskip}
0.95 &45.098003329505353    &4539479.1711295677    &5891663480815.9028    \\
     &45.098003329505\u{410}&4539479.171129567\u{5}&5891663480815.90\u{62}\\
\noalign{\smallskip}
0.98 &45.017931867960028    &4528085.2709957805    &5853723375274.4502    \\
     &45.0179318679\u{59945}&4528085.270995780\u{8}&5853723375274.45\u{31}\\
\noalign{\smallskip}
0.99 &45.004943101537128    &4526207.1227753328    &5847407841550.2810    \\
     &45.004943101537\u{080}&4526207.122775332\u{1}&5847407841550.28\u{52}\\
\noalign{\smallskip}
\multicolumn{1}{l}{$1$} &
\multicolumn{1}{c}{$45$} &
\multicolumn{1}{c}{$a/\sqrt2$} &
\multicolumn{1}{c}{$\frac18(\pi-2) a^2$} \\
\noalign{\smallskip}\hline\hline
\end{tabular}
\end{center}
\end{table}%
\begin{table}[tb]
\caption{\label{prolate-tab}
Analogous results to Table~\ref{oblate-tab}, but for prolate ellipsoids.
Here the values for $S_{12}$ for $-n \ge 0.98$ exceed $10^{17}\,\mathrm
m^2$ and these are displayed to the nearest whole square meter, i.e., 18
decimal digits.}
\begin{center}\let\u\underline
\begin{tabular}{>{$\scriptstyle}l<{$}>{$\scriptstyle}r<{$}>
{$\scriptstyle}r<{$}>{$\scriptstyle}r<{$}}
\hline\hline\noalign{\smallskip}
\multicolumn{1}{c}{$-n$} & \multicolumn{1}{c}{$\lambda_2\,({}^\circ)$} &
\multicolumn{1}{c}{$s_{12}\,(\mathrm{m})$} &
\multicolumn{1}{c}{$S_{12}\,(\mathrm{m}^2)$}
\\\noalign{\smallskip}\hline\noalign{\smallskip}
0.01 &91.288854749527201    &10205732.514416281    &33164247992795.212    \\
     &91.288854749527\u{189}&10205732.514416281    &33164247992795.21\u{1}\\
\noalign{\smallskip}
0.02 &92.610457370098315    &10362118.908653340    &34198331462649.689    \\
     &92.61045737009831\u{7}&10362118.9086533\u{39}&34198331462649.6\u{95}\\
\noalign{\smallskip}
0.05 &96.781576904333249    &10854904.416431548    &37558272805333.618    \\
     &96.7815769043332\u{35}&10854904.4164315\u{50}&37558272805333.6\u{25}\\
\noalign{\smallskip}
0.10 &104.48653831623701    &11762457.095994598    &44149951026541.638    \\
     &104.4865383162370\u{0}&11762457.095994\u{601}&44149951026541.6\u{48}\\
\noalign{\smallskip}
0.20 &123.32603446808286    &13970425.888241007    &62377755412860.708    \\
     &123.3260344680828\u{5}&13970425.88824100\u{2}&62377755412860.70\u{3}\\
\noalign{\smallskip}
0.40 &182.39739178709022    &20839831.771249872    &139011560027117.22    \\
     &182.3973917870902\u{1}&20839831.7712498\u{68}&139011560027117.2\u{8}\\
\noalign{\smallskip}
0.60 &304.70849870674712    &34975034.400175888    &391742895540755.96    \\
     &304.7084987067471\u{5}&34975034.4001758\u{69}&391742895540755.9\u{4}\\
\noalign{\smallskip}
0.90 &1428.1147116097373    &164323044.22719251    &8648966122417968.8    \\
     &1428.114711609737\u{9}&164323044.2271925\u{2}&86489661224179\u{73.0}\\
\noalign{\smallskip}
0.95 &2929.9802152369325    &337162015.21658508    &36412212193071484    \\
     &2929.9802152369\u{275}&337162015.21658\u{468}&364122121930714\u{24}\\
\noalign{\smallskip}
0.98 &7436.6985105821765    &855784235.25896030    &234584335066399637    \\
     &7436.6985105821\u{568}&855784235.2589\u{5941}&234584335066399\u{136}\\
\noalign{\smallskip}
0.99 &14948.252975667299    &1720188142.2370088    &947812505054464151    \\
     &14948.2529756672\u{32}&1720188142.23700\u{64}&94781250505446\u{5024}\\
\noalign{\smallskip}\hline\hline
\end{tabular}
\end{center}
\end{table}%
We start by considering the geodesic that starts on the equator, i.e.,
at its node $\phi_1 = 0$, with azimuth $\alpha_0 = \alpha_1
= \frac14\pi$ and compute the longitude at the vertex (the point of
maximum latitude), the distance to the vertex, and the area under this
geodesic segment.  We shall carry out this calculation with various
values of the third flattening $n$.  The arc length on the auxiliary
sphere, in this case, is $\sigma_{12} = \half\pi$, and, because
$\tan\alpha_0 = 1$, we have $\tan\phi_2 = (1+n)/(1-n)$.

The results are shown in Tables \ref{oblate-tab} and \ref{prolate-tab}
for oblate and prolate ellipsoids respectively.  For each $n$, the first
row shows the results with high precision arithmetic rounded to 17
decimal digits.  These allow independent verification of the algorithms.
The second row shows the results of carrying out the calculation in
double precision.  In all cases, the roundoff errors are very
small---the changes take place in no more than the last 5 bits of the
fraction of the binary representation; with $\abs n \le 0.9$,
the errors amount to at most 7 units in the last place in the binary
representation.

A more challenging test is presented with the data for the
administrative boundaries of Poland (this was kindly provided to me in
2013 by Paweł Pędzich of the Warsaw University of Technology).  This set
includes a polygonal representation of the boundary of Poland itself.
This consists of $67\,801$ edges with mean length $53\,\mathrm m$,
minimum length $62\,\mathrm{mm}$, and maximum length $9.7\,\mathrm{km}$;
the perimeter of the polygon is about $3\,600\,\mathrm{km}$ and its area
is about $313\,000\,\mathrm{km}^2$.  This is a good test for geodesic
algorithms as it represents a realistic ``large'' problem allowing the
overall accuracy and speed of the algorithms to be assessed.

In performing tests on ellipsoids with varying eccentricities, we need
to stipulate how the points are mapped onto the ellipsoid.  If we merely
keep the latitude fixed, then for extreme oblate, resp.~prolate,
ellipsoids, all the points end up concentrated near the equator,
resp.~the poles, making it difficult to interpret the results.  We
instead map the points keeping the authalic latitude fixed.  If, in
addition, we choose $a$ so that the area of the ellipsoid matches that
of the Earth, the area of the polygon is nearly constant as $n$ is
varied and we assess the errors by computing the difference between the
computed area and the true area.  For this exercise, we find the true
area using quadruple precision arithmetic.

\begin{figure}[tb]
\begin{center}% produced by poland.m
\includegraphics[scale=0.75]{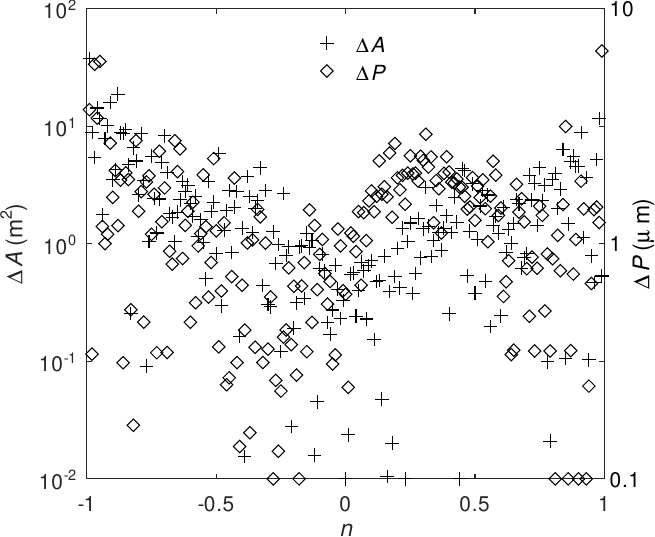}
\end{center}
\caption{\label{poland-err}
The absolute errors in the calculation of the area $\Delta A$ (plus
signs, scale on the left) and perimeter $\Delta P$ (diamonds, scale on
the right) of a polygonal model of Poland.  Here the model is mapped to
ellipsoids with various eccentricities (denoted by the third flattening
$n$) by preserving the value of the authalic latitude.}
\end{figure}%
The errors in the area and the perimeter are shown in
Fig.~\ref{poland-err}.  Generally speaking the errors in the area are
between $1\,\mathrm m^2$ and $10\,\mathrm m^2$ and the errors in the
perimeter are between $1\,\mu\mathrm m$ and $3\,\mu\mathrm m$.

The elapsed time for these calculations is approximated by
\[
\bigl(4.5 + N(40+\log_2N)/450\bigr)\,\mathrm{\mu s} \text{ per edge},
\]
where the first term represents the time to compute the perimeter of the
polygon, which involves solving the inverse geodesic problem for an
edge, and the second term is the additional time required to compute its
area.  Here $N$ is the number of terms used in the DST which is given in
Fig.~\ref{N-double}.  The time to compute the area begins to dominate
for $\abs n > 0.2$.  The area calculation could be made faster
if the DSTs were computed using the FFTW package instead of KISS FFT.

\section{Discussion}

In this paper, we have shown how to solve geodesic problems on an
ellipsoid of arbitrary eccentricity and an implementation of these
methods is available in GeographicLib.  The work naturally divides into
two parts:

The method of solving the standard direct and inverse problems was an
exercise in applying Carlson's algorithms for the evaluation of elliptic
integrals to the work of Legendre and Cayley.  These then replaced the
Taylor series expansions for the elliptic integrals used in {\it AG}.
In addition, the method for solving the inverse problem needed to be
adjusted to ensure convergence in all cases.

The evaluation of the area integral, Eq.(\ref{i4int}), is simplicity
itself: write the periodic integrand as a Fourier series, evaluate the
Fourier coefficients by trapezoidal quadrature, and then the integral of
the resulting series is trivial.  This is an accurate and fast method
because (1)~the Fourier series of an analytic function converges
rapidly, (2)~quadrature is very accurate when applied to the full period
of a periodic function, (3)~the FFT can be used to evaluate all the
coefficients efficiently.  This method will be useful in other
applications, e.g., to evaluate the Abelian integrals appearing in the
solution given by \citet{jacobi39} for geodesics on a triaxial
ellipsoid.

The algorithms have been tested for a wide range of eccentricities,
$\abs n \le 0.99$ or $1/199 \le b/a \le 199$; this encompasses
ellipsoids flatter than a pancake and as thin as spaghetti.  In principle,
the methods would work at more extreme eccentricities; however, it is
likely a more specialized approach would be better in these cases.
Despite this, the Taylor series method described in {\it AG} is
preferred for small flattening, $\abs f \le \frac1{50}$.  This is
somewhat more accurate and somewhat faster than the general method given
here.

The obvious domain of application for these algorithms is studying
planets and other celestial bodies with high eccentricity.  But, they
also play a role in terrestrial applications.
First of all, they allow the errors in the Taylor series method to be
measured accurately.  Secondly, because the flattening of terrestrial
ellipsoids is so small, a starting
approximation to the solution is given by spherical trigonometry and
this solution can be subsequently refined using ellipsoidal geodesics.
Unfortunately, any specifically ellipsoidal
behavior might be overlooked, precisely because the flattening is small.
Therefore, it behooves the developer to test the algorithms on more
eccentric ellipsoids, say, with $f = \frac15$, because then any
ellipsoid effects will be much more apparent.

\ifeprint\bibliography{geod}\else                  %%
\section*{Declarations}

\paragraph*{Data availability}
The data used to compute the area of Poland in Sec.~\ref{results} is
available upon request.
\paragraph*{Code availability}
An implementation of the algorithms given in this paper is available in
GeographicLib, version
2.1.2, available at
\url{https://github.com/geographiclib/geographiclib/releases/tag/r2.1.2}.
\paragraph*{Funding}
No funds, grants, or other support was received.
\paragraph*{Conflict of interest}
The author has no conflict of interest.
\fi                                                %%

\begin{appendix}

\section{Addenda to Karney (2013)}

\subsection{Multiple shortest geodesics} \label{multiple-shortest}

The shortest distance found by solving the inverse problem is
(obviously) uniquely defined.  However, in some special cases, multiple
azimuths yield the same shortest distance.  Here is a catalog of those
cases:
\begin{itemize}
\ifeprint\itemsep-0.4ex plus 0.5ex \fi %%
\item
$\phi_1 = -\phi_2$ (with neither point at a pole): If $\alpha_1
= \alpha_2$, the geodesic is unique.  Otherwise, there are two geodesics
and the second one is obtained by setting
$[\alpha_1,\alpha_2] \leftarrow [\alpha_2,\alpha_1]$,
$[M_{12},M_{21}] \leftarrow [M_{21},M_{12}]$, and $S_{12} \leftarrow
-S_{12}$.  (This occurs when $\lambda_{12}$ is close to $\pm180^\circ$
for oblate ellipsoids.)
\item
$\lambda_{12} = \pm 180^\circ$ (with neither point at a pole).  If
$\alpha_1 = 0^\circ$ or $\pm180^\circ$, the geodesic is unique.
Otherwise, there are two geodesics and the second one is obtained by
setting $[\alpha_1,\alpha_2] \leftarrow [-\alpha_1,-\alpha_2]$,
$S_{12} \leftarrow -S_{12}$.  (This occurs when $\phi_2$ is close to
$-\phi_1$ for prolate ellipsoids.)
\item
points 1 and 2 are at opposite poles: There are infinitely many
geodesics that can be generated by setting
$[\alpha_1,\alpha_2] \leftarrow [\alpha_1,\alpha_2] + [\delta,-\delta]$,
for arbitrary $\delta$.  (For spheres, this prescription applies when
points 1 and 2 are antipodal.)
\item
$s_{12} = 0$ (coincident points): There are infinitely many geodesics
that can be generated by setting $[\alpha_1,\alpha_2] \leftarrow
[\alpha_1,\alpha_2] + [\delta,\delta]$, for arbitrary $\delta$.
\end{itemize}
In cases where there are multiple shortest geodesics, GeographicLib
returns an arbitrary one.  The list given above allows the user (a)~to
determine whether there are other shortest geodesics and (b)~to generate
them.

\subsection{Refining the result from the reverted distance series}
\label{reverted-dist}
The 6th-order series given in {\it AG} provide solutions for the
geodesic problem which are accurate to roundoff for $\abs f \le
\frac1{100}$.  The least accurate of the series is the reverted series for
$\sigma$ in terms of the scaled distance $\tau$, Eqs.~(20) and (21) of
{\it AG}, which is used only in solving the direct problem.  (This
series also gives the reduced latitude $\beta$ in terms of the
rectifying latitude $\mu$.)  The accuracy is improved by using these
equations to give an initial approximation for $\sigma$ which is
followed by one step of Newton's method applied to Eq.~(\ref{s-int}),
with $\d s/\d\sigma = b\sqrt{1 + k^2 \sin^2\sigma}$.  With this change
(which need only be applied for $\abs f > \frac1{100}$), the 6th-order
series are accurate to roundoff for $\abs f \le \frac1{50}$.

\subsection{Non-equatorial geodesics} \label{equatorial-ends}
Care needs to be taken when solving the inverse problem for a
non-equatorial geodesic when both endpoints are on the equator.  If
$\phi_1 = \phi_2 = 0$, set $\phi_1 = -0$ when determining $\sigma_1$ and
$\omega_1$ using Eqs.~(11) and (12) of {\it AG}.  The library function
atan2 will use the sign of zero to determine the correct quadrant.  In
addition, $\alpha_1 = \half\pi$ should be treated as $\alpha_1
= \half\pi+\delta$ (e.g., by setting $\cos\alpha_1$ to some tiny
negative value $-\delta$); this breaks the degeneracy in the equation
for $\sigma_1$.

\subsection{An improved series for $A_2$}
The series for $A_2$, Eq.~(42) in {\it AG}, converges slightly faster if
we expand ${(1+\epsilon)}A_2$, instead of $A_2/{(1-\epsilon)}$.  The
resulting series is
\begin{equation}
\textstyle
A_2 = (1 + \epsilon)^{-1} \bigl(
1 - \frac{3}{4}\epsilon^2 - \frac{7}{64}\epsilon^4
- \frac{11}{256}\epsilon^6 + \ldots
\bigr).
\end{equation}

\subsection{Re-expansion of the area integral}
\label{areaseries}
Equations (63) of {\it AG} give coefficients $C_{4l}$ for the Fourier
series for the area integral.  These are Taylor series with $e'^2$ and
$k^2$ as small parameters.  The radius of convergence for this series is
$\abs{e'} = 1$; this means that the series diverges for oblate
ellipsoids with $a > \sqrt2b$.  This problem is remedied by
expanding, instead, in terms of
\begin{equation}\label{eps-def}
n = \frac{\sqrt{1+e'^2}-1}{\sqrt{1+e'^2}+1},\quad
\epsilon = \frac{\sqrt{1+k^2}-1}{\sqrt{1+k^2}+1},
\end{equation}
the same small parameters used by Helmert for the Taylor series
expansion of the longitude integral, Eq.~(\ref{lambda-int-a}).  The
radius of convergence of the resulting series is $\abs n = 1$
covering all eccentricities.  The Fourier coefficients can then be
written as
\begin{align}
C_{40} &\textstyle= \bigl(\frac{2}{3} - \frac{4}{15} n + \frac{8}{105} n^2 +
\frac{4}{315} n^3 + \frac{16}{3465} n^4 + \frac{20}{9009} n^5\bigr)\notag\\
&\textstyle\quad{}- \bigl(\frac{1}{5} - \frac{16}{35} n +
\frac{32}{105} n^2 - \frac{16}{385} n^3 -
\frac{64}{15015} n^4\bigr) \epsilon\notag\\
&\textstyle\quad{}- \bigl(\frac{2}{105} + \frac{32}{315} n -
\frac{1088}{3465} n^2 + \frac{1184}{5005} n^3\bigr) \epsilon^2\notag\\
&\textstyle\quad{}+ \bigl(\frac{11}{315} - \frac{368}{3465} n -
\frac{32}{6435} n^2\bigr) \epsilon^3\notag\\
&\textstyle\quad{}+ \bigl(\frac{4}{1155} +
\frac{1088}{45045} n\bigr) \epsilon^4
+ \frac{97}{15015} \epsilon^5 + \ldots,\displaybreak[0]\notag\\
C_{41} &\textstyle=  \bigl(\frac{1}{45} - \frac{16}{315} n +
\frac{32}{945} n^2 - \frac{16}{3465} n^3 -
\frac{64}{135135} n^4\bigr) \epsilon\notag\\
&\textstyle\quad{}- \bigl(\frac{2}{105} - \frac{64}{945} n +
\frac{128}{1485} n^2 - \frac{1984}{45045} n^3\bigr) \epsilon^2\notag\\
&\textstyle\quad{}- \bigl(\frac{1}{105} - \frac{16}{2079} n -
\frac{5792}{135135} n^2\bigr) \epsilon^3\notag\\
&\textstyle\quad{}+ \bigl(\frac{4}{1155} -
\frac{2944}{135135} n\bigr) \epsilon^4
+ \frac{1}{9009} \epsilon^5 + \ldots,\displaybreak[0]\notag\\
C_{42} &\textstyle=  \bigl(\frac{4}{525} - \frac{32}{1575} n +
\frac{64}{3465} n^2 - \frac{32}{5005} n^3\bigr) \epsilon^2\notag\\
&\textstyle\quad{}- \bigl(\frac{8}{1575} - \frac{128}{5775} n +
\frac{256}{6825} n^2\bigr) \epsilon^3\notag\\
&\textstyle\quad{}- \bigl(\frac{8}{1925} -
\frac{1856}{225225} n\bigr) \epsilon^4
+  \frac{8}{10725} \epsilon^5 + \ldots,\displaybreak[0]\notag\\
C_{43} &\textstyle=  \bigl(\frac{8}{2205} - \frac{256}{24255} n +
\frac{512}{45045} n^2\bigr) \epsilon^3\notag\\
&\textstyle\quad{}- \bigl(\frac{16}{8085} -
\frac{1024}{105105} n\bigr) \epsilon^4
- \frac{136}{63063} \epsilon^5 + \ldots,\displaybreak[0]\notag\\
C_{44} &\textstyle=  \bigl(\frac{64}{31185} -
\frac{512}{81081} n\bigr) \epsilon^4
- \frac{128}{135135} \epsilon^5 + \ldots,\displaybreak[0]\notag\\
C_{45} &\textstyle=  \frac{128}{99099} \epsilon^5 + \ldots.
\label{c4-series}
\end{align}
At the order shown, this series gives full double precision accuracy for
the area for $\abs f \le \frac1{50}$.

\section{Comparison with Nowak et al.\ (2022)}

\citet[here called {\it NNdC}]{nowak22} also consider the problem of
geodesics on an arbitrary ellipsoid.  Their starting point for
evaluating the integrals for the distance and longitude is expanding
$\sqrt{1-e^2}$ as a Taylor series.  So, as a preliminary matter, their
methods fail to converge for $\abs{e^2} > 1$, i.e., for prolate
ellipsoids with $b > \sqrt2a$.  Furthermore, their results for the
integrals are unnecessarily complicated given that these can be put in
closed form in terms of elliptic integrals and evaluated
using \citet{carlson95}; see Sec.~\ref{elliptic-sec}.

Therefore, let us concentrate on their series for computing the area.
Their expression for the area, Eqs.~(72), is the same as our
Eqs.~(\ref{Sdef}) with the substitution
\begin{equation}
p(\sigma) = \frac{a^2}{c^2}
\sin2\alpha\Ar(C^2, \rho^2),
\end{equation}
where
\begin{equation}
C^2 = \sin^2\alpha_0,\quad
\rho^2 = 1- \cos^2\alpha_0 \sin^2\sigma.
\end{equation}
\begin{table}[tb]
\caption{\label{nowak-tab}
The minimum number of terms $N_{\mathrm{NNdC}}$ required for full double
precision accuracy for the area using the algorithm of \citet{nowak22}.
The results are given for the computation of the area under a geodesic
segment between its node and its vertex with $\alpha_0=\frac14\pi$.
(This is the test case considered in Tables~\ref{oblate-tab}
and \ref{prolate-tab}.)  The method diverges for $e^2 < -1$ or
$n \lesssim -0.1716$.  The corresponding number of terms
needed using the DST method $N_{\mathrm{DST}}$ is given in the last column.}
\begin{center}
\begin{tabular}{@{\extracolsep{1.5em}}>{$}r<{$}>{$}r<{$}>{$}r<{$}}
\hline\hline\noalign{\smallskip}
\multicolumn{1}{c}{$n$} &
\multicolumn{1}{c}{$N_{\mathrm{NNdC}}$} &
\multicolumn{1}{c}{$N_{\mathrm{DST}}$}
\\\noalign{\smallskip}\hline\noalign{\smallskip}
  -0.18& \text{diverges}&  16\\
  -0.17&    1241        &  15\\
  -0.15&     118        &  14\\
  -0.10&      35        &  12\\
  -0.05&      18        &   9\\
\pm0.01&       9        &   6\\
\pm0.02&      11        &   7\\
   0.05&      16        &   9\\
   0.10&      24        &  12\\
   0.20&      43        &  16\\
   0.30&      72        &  21\\
   0.40&     117        &  27\\
   0.60&     337        &  44\\
   0.80&    1522        &  88\\
   0.90&    5873        & 160\\
   0.95&   20992        & 288\\
   0.99& \mathord>100000& 912\\
\noalign{\smallskip}\hline\hline
\end{tabular}
\end{center}
\end{table}%
The function $\Ar(C^2, \rho^2)$ is, in turn, given by an infinite sum,
their Eq.~(69).  For the case considered in Tables~\ref{oblate-tab}
and \ref{prolate-tab}, namely computing the area under a geodesic
segment between its node and its vertex with $\alpha_0=\frac14\pi$, the
two methods yield the same results provided that $\abs{e^2} < 1$;
this serves to validate both approaches.  We can therefore include
enough terms in the infinite sum to limit the errors in the area to
$2^{-53}c^2$, the same criterion we used in selecting the number of
points to use for the DST.  Table~\ref{nowak-tab} compares this number,
$N_{\mathrm{NNdC}}$, to the corresponding number for the DST,
$N_{\mathrm{DST}}$, from Fig.~\ref{N-double}.  We can make a few points
here:
\begin{enumerate}
\ifeprint\itemsep-0.4ex plus 0.3ex \fi %%
\item $N_{\mathrm{NNdC}}$ is consistently
greater than $N_{\mathrm{DST}}$, in some cases, by a large factor.
\item The value $N_{\mathrm{DST}}$ is selected to meet the accuracy
threshold for {\it all} geodesics on an ellipsoid with the given $n$,
while, here, I determined $N_{\mathrm{NNdC}}$ only for one particular
value $\alpha_0 = \frac14\pi$.
\item The computational cost for evaluating the area
integral with the method of {\it NNdC} scales as
$O(N_{\mathrm{NNdC}}^2)$ which is more expensive than for the DST
method, $O(N_{\mathrm{DST}}\log N_{\mathrm{DST}})$.
\item The results in Table~\ref{nowak-tab} were found using high
precision arithmetic and so reflect only the truncation errors in both
methods.  {\it NNdC} have not provided an implementation of their
method, so it's not possible to compare either the roundoff errors or
the timing; however, considering points 1 and 3, it's likely that their
method is less accurate and slower than the DST method.
\end{enumerate}

\end{appendix}

\ifeprint\else       %%
\bibliography{geod}
\fi                  %%

\end{document}